\documentclass{article}
\usepackage[cp1251]{inputenc}
\usepackage[english,russian]{babel}
\inputencoding{cp1251}
\author{A.\,A.\,Stanislavsky\thanks{alexstan@ri.kharkov.ua}\\
\\ Institute of Radio Astronomy, Ukrainian National Academy of
Sciences,\\ 61002 Kharkov, Ukraine}
\title{\bf Beam propagation in a Randomly Inhomogeneous Medium}
\date{\small Published in {\it Journal of Experimental and Theoretical
Physics} {\bf 98}(4), 2004, pp. 705-706}
\begin{document}
\large\tolerance8000\hbadness10000\emergencystretch3mm \maketitle
\maketitle

\begin{abstract}
An integro-differential equation describing the angular
distribution of beams is analyzed for a medium with random
inhomogeneities. Beams are trapped because inhomogeneities give
rise to wave localization at random locations and random times.
The expressions obtained for the mean square deviation from the
initial direction of beam propagation generalize the ``3/2 law''.
\end{abstract}


Fluctuations of the beam propagation direction in a randomly
inhomogeneous medium are frequently observed in nature.  Examples
include random refraction of radio waves in the ionosphere and
solar corona, stellar scintillation due to atmospheric
inhomogeneities, and other phenomena. The propagation of a beam
(of light, radio waves, or sound) in such media can be described
as a normal diffusion process ~\cite{bib:R, bib:Ch1}. One
extraordinary property predicted for random media -- and later
revealed -- is the Anderson localization~\cite{bib:A}. It brings
normal diffusion to a complete halt. In the context of wave
propagation in a random medium, the Anderson localization is
caused by interference of waves resulting from multiple scattering
~\cite{bib:W}. When two waves propagating in opposite directions
along a closed path are in phase, the resultant wave is more
likely to return to the starting point than propagate in other
directions. The properties of a randomly inhomogeneous medium vary
not only from point to point, but also with time. Consequently,
localization may take place both at random locations and at random
times. Random localization affects diffusive light propagation in
a random medium. An approach to description of this effect is
developed in this paper.

Suppose that the medium is statistically homogeneous and
isotropic. Then, a beam propagating through the medium is
deflected at random. Localization implies that the beam is trapped
in some region. Since the trapped beam returns to the point where
it was trapped, its propagation is ``frozen'' for some time. After
that, a randomly deflected beam leaves the legion and propagates
further until it is trapped in another region (or at a point) and
the localization cycle repeats. The randomly winding beam path due
to inhomogeneities is responsible for the random refraction
analyzed in this study.

The angle $\theta$ of deviation of a beam from its initial
direction is characterized by a probability density
$W_\alpha(\theta, \sigma)$, where $\sigma$ is the path travelled
by the beam. Let us derive an integro-differential equation for
the probability density. In contrast to rotational Brownian
motion, the random walks analyzed here consist of random angle
jumps $\Delta\theta_i$ at points separated by segments of random
length $\Delta\sigma_i$. Exact knowledge of the distributions of
these random variables is not required. It is sufficient to assume
that the angle jumps are independent random variables belonging
the domain of attraction of Gaussian probability distribution. The
random segment lengths $\Delta\sigma_i$ are also identically
distributed independent random variables, with stable distribution
is characterized by an index $\alpha$. Since $\Delta\sigma_i$ is a
nonnegative quantity, this distribution is totally asymmetric, and
$0<\alpha\leq1$. Recall that a random variable characterized by a
probability distribution $f(x)$, of this kind is described by the
Laplace transform
\begin{displaymath}
\phi(s)=\int^\infty_{0}\exp\{-sx\}\,df(x)=\exp\{-(Ax)^\alpha\},
\end{displaymath}
where $x\geq 0$ and $A>0$~\cite{bib:Z}. The total path length is
the sum of all $\Delta\sigma_i$. Both $\Delta\sigma_i$ and
$\Delta\theta_i$ are Markov processes. However, since the former
is the master process with respect to the latter, the resultant
process may not preserve the Markov property~\cite{bib:F}. On
account of convergence of distributions we can definitely pass
from the discrete model to a continuous limit~\cite{bib:S}. This
leads to the diffusion equation
\begin{displaymath}
W_\alpha(\theta,\sigma)-W_\alpha(\theta,0)=
\int_0^\sigma\frac{D}{\Gamma(\alpha)\sin\theta}
\frac{\partial}{\partial\theta}\left(\sin\theta\frac{\partial
W_\alpha(\theta,\sigma')}{\partial\theta}\right)
(\sigma-\sigma')^{\alpha-1}\,d\sigma'\,,
\end{displaymath}
where $D$ is a diffusion coefficient and $\Gamma(x)$ is the gamma
function. The solution to this equation can be expressed as an
integral transform of the probability density associated with
rotational Brownian motion:
\begin{displaymath}
W_\alpha(\theta,\sigma)=\int^\infty_0F_\alpha(z)\,
W_1(\theta,\sigma^\alpha z)\,dz\,,
\end{displaymath}
where
\begin{displaymath}
F_\alpha(z)=\sum_{k=0}^\infty\frac{(-z)^k}{k!
\Gamma(1-\alpha-k\alpha)}.
\end{displaymath}
The diffusion equation yields the mean
\begin{displaymath}
\overline{\cos\theta}=E_\alpha(-2D\sigma^\alpha)\,,
\end{displaymath}
where
\begin{displaymath}
E_\alpha(-x)=\sum_{n=0}^\infty(-x)^n /\Gamma(1+n\alpha)
\end{displaymath}
is the Mittag-Leffler function. At large $\sigma$, all beam
directions are equiprobable. However, in contrast to normal
diffusion ($\alpha=1$), a beam has to travel a longer path
$\sigma$ to reach this state. This process is somewhat analogous
to ``superslow'' relaxation.

Following the method developed in~\cite{bib:Ch2}, one can find the
mean square of the distance $r$ from the starting point to the
observation point reached by the beam that has travelled an
intricate path of length $\sigma$ through the medium:
\begin{equation}
\overline{r^2}=\frac{\sigma^\alpha}{D\Gamma(\alpha+1)}-\frac{1}
{2D^2}\Bigl(1-E_\alpha(-2D\sigma^\alpha)\Bigr).\label{eq:1}
\end{equation}
If $D\sigma\ll 1$, then
\begin{equation}
\overline{r^2}\approx
2\sigma^{2\alpha}\left(\frac{1}{\Gamma(2\alpha+1)}-\frac
{2D\sigma^\alpha}{\Gamma(3\alpha+1)}\right).\label{eq:2}
\end{equation}
If the  $z$ axis of a polar coordinate system is aligned with the
initial beam direction, then the mean square of the distance
passed by the beam along this axis is given by the formula
\begin{equation}
\overline{z^2}=\frac{1}{3D}\left[\frac{\sigma^\alpha}{\Gamma(\alpha+1)}
-\frac{1}{6D}\Bigl(1-E_\alpha(-6D\sigma^\alpha)\Bigr)\right].
\label{eq:3}
\end{equation}
If $D\sigma$ is small, then
\begin{equation}
\overline{z^2}\approx
2\sigma^{2\alpha}\left[\frac{1}{\Gamma(2\alpha+1)}-\frac
{6D\sigma^\alpha}{\Gamma(3\alpha+1)}\right].\label{eq:4}
\end{equation}
Now, the mean square deviation of the beam from its initial
direction can be calculated by combining (\ref{eq:1})
with~(\ref{eq:3}):
\begin{equation}
\overline{\rho^2}=\overline{r^2}-\overline{z^2}=
\frac{2\sigma^\alpha}{3D\Gamma(\alpha+1)}-\frac{1}{2D^2} \Bigl(1-
E_\alpha(-2D\sigma^\alpha)\Bigr)+\frac{1}{18D^2}
\Bigl(1-E_\alpha(-6D\sigma^\alpha)\Bigr). \label{eq:5}
\end{equation}
If $D\sigma$ is small, then a generalized ``3/2 law'' is obtained
~\cite{bib:Ch2}:
\begin{equation}
\sqrt{\overline{\rho^2}}\approx\frac{2\sqrt{2}}{\sqrt{\Gamma(3\alpha+1)}}
D^{1/2}\sigma^{3\alpha/2}. \label{eq:6}
\end{equation}
The mean squares given by ~(\ref{eq:1}), (\ref{eq:3})
and~(\ref{eq:5}) increase as $\sigma^\alpha$ at large $\sigma$.
The case of $\alpha=1$ corresponds to normal diffusion without
wave localization. Thus, the approach developed here embraces
classical results of the theory of beam propagation medium in a
randomly inhomogeneous medium~\cite{bib:R, bib:Ch1}.

Finally,   it should  be  recalled that experimental deviations
from the ``3/2 law'' (more precisely, from an exponent of 3/2 in
the classical power law) were mentioned in~\cite{bib:K}. However,
they were attributed to systematic measurement errors, probably
because of the lack of plausible interpretation. This problem can
be revisited in view of the results obtained in this study.
Moreover, new accurate experimental studies of beam propagation in
appropriate randomly inhomogeneous media would be extremely useful
for verifying the model proposed here.


\begin{thebibliography}{2003}

\bibitem{bib:R}
S.~M.~Rytov, {\it Introduction to Statistical Radiophysics}
(Nauka, Moscow, 1966).

\bibitem{bib:Ch1}
L.~A.~Chernov, {\it Wave Propagation in a Random Medium} (Nauka,
Moscow, 1975; McGraw-Hill, New York, 1960).

\bibitem{bib:A}
P.\,W. Anderson, Phys. Rev. {\bf 109}, 1492 (1958).

\bibitem{bib:W}
D.\,S. Wiersma, P. Bartolini, A. Lagendijk, R. Righini, Nature
{\bf 390}, 671 (1997).

\bibitem{bib:Z}
V.~M.~Zolotarev, {\it One-Dimensional Stable Distributions}
(Nauka, Moscow, 1983; Am. Math. Soc., Providence, RI, 1986).

\bibitem{bib:F}
W.~Feller, {\it An Introduction to Probability Theory and Its
Applications}, 3rd ed. (Wiley, New York, 1967; Mir, Moscow, 1964).

\bibitem{bib:S}
A.\,A. Stanislavsky, Phys. Scripta {\bf 67}, 265 (2003).

\bibitem{bib:Ch2}
L.~A.~Chernov, Zh. Eksp. Teor. Fiz. {\bf 24}, 210 (1953).

\bibitem{bib:K}
I.~G.~Kolchinskii, Astron. Zh.  {\bf 29}, 350 (1952).

\end{thebibliography}
\end{document}